\documentclass[aoas,preprint]{imsart}

\RequirePackage[OT1]{fontenc}
\RequirePackage{amsthm,amsmath,amssymb,subcaption,booktabs,graphicx}
\RequirePackage{algcompatible, algorithm}
\RequirePackage[authoryear]{natbib}
\RequirePackage[colorlinks,citecolor=blue,urlcolor=blue]{hyperref}
\RequirePackage[table]{xcolor}
\RequirePackage{tikz}
\usetikzlibrary{arrows,snakes,backgrounds}
\usepackage{enumitem}
\usepackage{subcaption}
\usepackage{amsbsy}%letras gregas em negrito

\arxiv{arXiv:0000.0000}

\startlocaldefs
\numberwithin{equation}{section}
\theoremstyle{plain}
\newtheorem{thm}{Theorem}[section]

\newtheorem{proposition}{Proposition}
\theoremstyle{definition}
\newtheorem{definition}{Definition}

\newtheorem{remark}{Remark}

\def\R{{\mathbb R}}

\def\Z{\mathbb Z}
\def\N{\mathbb{N}}
\def\p{^{\prime}}

\endlocaldefs

\begin{document}

\begin{frontmatter}
  \title{Context tree selection for  functional data} 
    \runtitle{Context tree selection for  functional data}
\begin{aug}
\author{\fnms{A.} \snm{Duarte}\thanksref{t2,m1}\ead[label=e1]{alineduarte@usp.br}},
\author{\fnms{R.} \snm{Fraiman}\thanksref{m2}\ead[label=e2]{rfraiman@cmat.edu.uy}},
\author{\fnms{A.} \snm{Galves}\thanksref{t3,m1}\ead[label=e3]{galves@usp.br}},
\author{\fnms{G.} \snm{Ost}\thanksref{t4,m1}\ead[label=e4]{guilhermeost@gmail.com}}
\and
\author{\fnms{C. D.} \snm{Vargas}\thanksref{t5,m3}
\ead[label=e5]{cdvargas@biof.ufrj.br}
}

\thankstext{t1}{December 18, 2017}
\thankstext{t2}{Fully and successively supported  by CNPq and FAPESP fellowships (grants  201696/2015-0 and 2016/17791-9).}
\thankstext{t3}{Partially supported  by CNPq fellowship grant 311 719/2016-3.}
\thankstext{t4}{Fully and successively supported by CNPq and FAPESP fellowships (grants  201572/2015-0 and 2016/17789-4).}
\thankstext{t5}{Partially supported by CNPq fellowship grant 306 817/2014-4.}
\runauthor{Duarte et al.}

\affiliation{Universidade de S\~ao Paulo\thanksmark{m1}, Universidad de la Rep\'ublica Uruguay\thanksmark{m2} and Universidade Federal do Rio de Janeiro \thanksmark{m3}}

\address{A. Duarte\\ A. Galves\\ G. Ost\\
Instituto de Matem\'atica e Estat\'istica\\
Universidade de S\~ao Paulo\\
Brazil\\
\printead{e1}\\
\printead{e3}\\
\printead{e4}
}

\address{R. Fraiman\\ 
Centro de Matem\'atica, Facultad de Ciencias\\
Universidad de la Rep\'ublica Uruguay\\
Urugay\\
\printead{e2}\\
}

\address{C. Vargas\\ 
Laborat\'orio de Neurobiologia II, Instituto de Biof\'isica Carlos Chagas Filho\\
Universidade Federal do Rio de Janeiro\\
Brazil\\
\printead{e5}\\
}
\end{aug}

\begin{abstract}

\begin{center}
{\it Dedicated to Jorma Rissanen on his $85^{th}$ birthday}
\end{center}

  It has been repeatedly conjectured that the brain retrieves
  statistical regularities from stimuli. Here we present a new
  statistical approach allowing to address this conjecture. This
  approach is based on a new class of stochastic processes driven by
  chains with memory of variable length. It leads to a new
  experimental protocol in which sequences of auditory stimuli
  generated by a stochastic chain are presented to volunteers while
  electroencephalographic (EEG) data is recorded from their scalp. A
  new statistical model selection procedure for functional data is
  introduced and proved to be consistent. Applied to samples of EEG
  data collected using our experimental protocol it produces results
  supporting the conjecture that the brain effectively identifies the
  structure of the chain generating the sequence of stimuli.

\end{abstract}



\end{frontmatter}

\section{Introduction}
\label{sec:intro}

Consider the following experimental situation.  A listener is exposed to
a sequence of auditory stimuli, generated by a stochastic
chain, while electroencephalographic (EEG) signals are recorded from
his scalp.  Starting from \cite{VonHelmholtz:67}, a classical conjecture in neurobiology claims that the
listener's brain automatically identifies statistical regularities in
the sequence of stimuli (see for instance \cite{Garrido:13} and
\cite{Wacongne:12}).  If this is the case, then a signature of the
stochastic chain generating the stimuli should somehow be encoded in
the brain activity. The question is whether this signature can be
identified in the EEG data recorded during the experiment. The goal of
this paper is to present a new statistical framework in which this
conjecture can be formally addressed and rigorously tested.

To model the relationship between the random chain of auditory stimuli
and the corresponding EEG data we introduce a new class
of stochastic processes. A process in this class has two
components. The first one is a stochastic chain taking values in the
set of auditory units. The second one is a sequence of functions
corresponding to the sequence of EEG chunks recorded during the exposure
of the successive auditory stimuli.

At this point we should decide which type of dependence from the past 
characterizes the sequence of auditory stimuli. Are the auditory units 
independent random variables? Do they constitute a Markov chain?
Besides modeling the chain of auditory units, we must also discuss how
to express  the relationship between the chain of stimuli and  the
sequence of EEG chunks.

Hidden Markov Models (HMM) have been used extensively in this type of
situation. A remarkable example is the work done by
Rabiner and co-authors to model natural language phonetics (see for
instance (\cite{Rabiner:1993}). Applied to our
experimental situation, this would mean modeling the sequence of
auditory units as a Markov chain in which each step would depend on a
finite fixed number of past units. Furthermore in the HMM framework
the law of each chunk of EEG would depend only on the auditory
stimulus presented during its recording.

It turns out that this last assumption is clearly insufficient to account 
for the well established phenomenon that the law 
of the EEG may depend not only on the value of the corresponding stimulus, 
but also on the fact that it has appeared or not in an expected situation 
(see for instance \cite{Naatanen:78} and the revision paper \cite{Naatanem:2005}).  
Modeling this type of experimental situation requires to assume that the 
law of the EEG chunk at each step depends on the smallest string of past stimuli 
which contains all the relevant information to predict the next stimulus in the sequence.

Stochastic chains with memory of variable length offer a natural
framework to present this kind of dependence from the past. Introduced
by \cite{Rissanen:83}, as a universal system for data compression,
they became known in the statistics community through
\cite{buhlmann99} in which they appear with the name of Variable
Length Markov Chains (VLMC). 
As an example of application,
they have been used by \cite{galves:12} to characterize rhythmic patterns in
linguistic data.

In his seminal paper \cite{Rissanen:83}, Rissanen observed that in many real life
stochastic chains the dependence from the past has not a fixed
length. Instead, it changes at each step as a function of the past
itself. He called a {\sl context} the smallest final string of past symbols
containing all the information required to predict the next
symbol.  The set of all contexts define a partition of the past and
can be represented by a rooted and labeled oriented tree. For this
reason many authors call  stochastic chains with memory of variable
length {\sl context tree models}. We shall adopt this terminology here.

The present paper discusses how to identify in the EEG data the signature 
of the chain generating the stimuli. This is a problem of statistical model 
selection for functional data which is, in general, a difficult and
unsolved issue. However the context
tree approach adopted here makes the question treatable. This is
done by the introduction of a new model selection procedure for
functional data driven by a context tree model. This procedure is
proved to be consistent. 

This article is organized as follows. In Section
\ref{sec:PresentationOfOurApproach} we present an informal overview of
our approach. In Section \ref{sec:HCTM} we introduce the notation, recall what is a {\sl context tree model}
and introduce the new class of {\sl stochastic processes driven by a context tree model}.  Our new procedure for
statistical model selection is presented in Section
\ref{sec.HCTMselection} together with Theorem \ref{thm:2} on the
consistency of the model selection procedure.  
Finally, in Section \ref{sec:case_study} we present a simulation study and also analyze a EEG dataset 
collected using our experimental protocol. The proofs of the theorems and propositions
are presented in the Appendices.

\section{Informal presentation of our approach}
\label{sec:PresentationOfOurApproach}
In our experimental protocol volunteers are exposed to sequences of
auditory stimuli generated by context tree models while EEG signals
are recorded.  
The auditory units used as stimuli  are
either {\it strong beats}, {\it weak beats} or {\it silent units}, represented by symbols $2, 1$ and $0$ respectively. 

The way the sequence of auditory units was generated can  be informally described as follows.
Start with a deterministic sequence, either

$$ 2 \ 1 \ 1\ 2 \ 1 \ 1 \ 2 \ 1 \ 1 \ 2 \ 1 \ 1 \ 2 \ldots .$$

or

$$2 \ 1 \ 0\ 1 \ 2 \ 1 \ 0 \ 1 \ 2 \ 1\ 0\ 1\ 2 \ldots$$

Then replace each weak beat (symbol $1$) by a silent unit (symbol $0$)
with probability $\epsilon$ in an independent way. 

An example of a
sequence produced by this procedure acting on the first (respectively second) basic sequence would be
$$
2 \ 1 \ 1\ 2 \ 0 \ 1 \ 2 \ 1 \ 1 \ 2 \ 0 \ 0 \ 2\ldots .
$$
and
$$
2 \ 1 \ 0\ 1 \ 2 \ 0 \ 0 \ 1 \ 2 \ 1\ 0\ 0\ 2 \ldots .
$$
Let us call {\em Ternary} and {\em Quaternary} respectively 
the stochastic chains generated in this way. In the
sequel these chains will be respectively denoted by the
symbols $(X^{ter}_0, X^{ter}_1, X^{ter}_2, \ldots)$  and
$(X^{qua}_0, X^{qua}_1, X^{qua}_2, \ldots)$.

Both stochastic chains can be generated step by step by an algorithm using only information from the past. We impose to the algorithm the condition that it uses, at each step, the shortest string of past symbols necessary to generate the next symbol.

In the case of the Ternary chain, this algorithm can be described as follows. To generate $X^{ter}_n$, given the past 
$X^{ter}_{n-1},X^{ter}_{n-2},\ldots$, we first look to the last symbol $X^{ter}_{n-1}$.

\begin{itemize}
\item If $X^{ter}_{n-1}=2$, then  
$$
X^{ter}_n=\left\lbrace\begin{array}{ll}
1,& \mbox{with probability} \ 1- \epsilon,\\
0, & \mbox{with probability} \ \epsilon.\
\end{array}\right.
$$
\item If $X^{ter}_{n-1}=1$ or $X^{ter}_{n-1}=0$, then we need to go back one more step, 
\begin{itemize}
\item[$\centerdot$] if $X^{ter}_{n-2}=2$, then
$$
X^{ter}_n=\left\lbrace\begin{array}{ll}
1,& \mbox{with probability} \ 1- \epsilon,\\
0, & \mbox{with probability} \ \epsilon;\
\end{array}\right.
$$ 
\item[$\centerdot$]  if $X^{ter}_{n-2}=1$ or $X^{ter}_{n-2}=0$, then $X^{ter}_n=2$ with probability $1$.
\end{itemize}
\end{itemize}

Similarly, the algorithm generating the Quaternary chain can be
described as follows. To generate $X^{qua}_n$, given the past $X^{qua}_{n-1},X^{qua}_{n-2},\ldots$, we first look to the last symbol $X^{qua}_{n-1}$.
\begin{itemize}
\item If $X^{qua}_{n-1}=2$, then  
$$
X^{qua}_n=\left\lbrace\begin{array}{ll}
1,& \mbox{with probability} \ 1- \epsilon,\\
0, & \mbox{with probability} \ \epsilon.\
\end{array}\right.
$$
\item If $X^{qua}_{n-1}=1$, then we need to go back one more step, 
\begin{itemize}
\item[$\centerdot$] if $X^{qua}_{n-2}=2$, then $X^{qua}_n=0$, with probability $1$. 
\item[$\centerdot$]  if $X^{qua}_{n-2}=0$, then $X^{qua}_n=2$, with probability $1$.
\end{itemize}
\item If $X^{qua}_{n-1}=0$, then we need to go back one more step,
\begin{itemize}
\item[$\centerdot$] if $X^{qua}_{n-2}=2$, then $X^{qua}_n=0$, with probability $1$. 
\item[$\centerdot$]  if $X^{qua}_{n-2}=1$, then
$$
X^{qua}_n=\left\lbrace\begin{array}{ll}
1,& \mbox{with probability} \ 1- \epsilon,\\
0, & \mbox{with probability} \ \epsilon;\
\end{array}\right.
$$ 
\item[$\centerdot$]  If $X^{qua}_{n-2}=0$, then we need to go back one more step,
\begin{itemize}
\item[$\centerdot$] if $X^{qua}_{n-3}=2$, then
$$
X^{qua}_n=\left\lbrace\begin{array}{ll}
1,& \mbox{with probability} \ 1- \epsilon,\\
0, & \mbox{with probability} \ \epsilon;\
\end{array}\right.
$$  
\item[$\centerdot$]  if $X^{qua}_{n-3}=1$ or $X^{qua}_{n-3}=0$, then $X^{qua}_n=2$, with probability $1$. 
\end{itemize}
\end{itemize}
\end{itemize}

The algorithms described above are characterized by two elements. 
The first one is a  partition of the set of all possible sequences of past units. 
In the case of the Ternary chain this partition is represented by the set 
 $$
 \tau_{ter}=\{00, 10, 20, 2, 01, 11, 21, 2  \}.
 $$
In this partition the string $00$ represents the set
 of all strings ending by the
 ordered pair $(0,0)$; $10$ represents the set of all strings ending by the
 ordered pair $(1,0)$, $\ldots$ and finally the symbol $2$ represents
 the set of all strings ending by $2$. 
 Following \cite{Rissanen:83}, let us call \textit{context} any element of this partition.

For instance, if
$$
\ldots,X^{ter}_{n-3}=1, X^{ter}_{n-2}=2, X^{ter}_{n-1}=0, X^{ter}_{n}=1.
$$ 
the context associated to this past sequence is $01$.

In the case of the Quaternary chain this partition is represented by the set 
 $$
 \tau_{qua}=\{000, 100, 200, 10, 20, 01, 21, 2\}.
 $$

\begin{figure}[h!]
\centering
\begin{tikzpicture}[thick,scale=0.8]
     \tikzstyle{level 1}=[level distance=1.5cm, sibling distance=2cm]
     \tikzstyle{level 2}=[level distance=1.3cm, sibling distance=0.7cm]
      \tikzstyle{level 3}=[level distance=1.3cm, sibling distance=0.7cm]
      \tikzstyle{invisible} = [rectangle, node distance=0.5cm]
%    \tikzstyle{every node}=[circle,draw]
    \coordinate
        child{{}
        	child { {}
        		child {[fill] circle (2.5pt) node(000){}} 
        		child {[fill] circle (2.5pt) node(100){}} 
        		child {[fill] circle (2.5pt) node(200){}} 
        		}
       		child {[fill] circle (2.5pt) node(10) {} }
        	child {[fill] circle (2.5pt) node(20) {} }		        	
        	}
        child{{} 
		child {[fill] circle (2.5pt) node(01){} }
        	child {[fill] circle (2.5pt) node(21){} }        
        }
        child{[fill] circle (2.5pt)  node(2){} }
    ;
    \node [ invisible, below of=000 ]{000};
    \node [ invisible, below of=100 ]{100};
    \node [ invisible, below of=200 ]{200};
    \node [ invisible, below of=10 ]{10};
    \node [ invisible, below of=20 ]{20};
    \node [ invisible, below of=01 ]{01};
    \node [ invisible, below of=21 ]{21};
    \node [ invisible, below of=2 ]{2};
\end{tikzpicture}
\hspace{30pt}
\raisebox{1cm}{
\begin{tikzpicture}[thick,scale=0.75]%negrito
     \tikzstyle{invisible} = [rectangle, node distance=0.5cm]
     \tikzstyle{level 1}=[level distance=1.5cm, sibling distance=2.2cm,]
     \tikzstyle{level 2}=[level distance=1.3cm, sibling distance=0.7cm]
    % \tikzstyle{every node}=[circle,draw]
    \coordinate
        child{{}
        	child {[fill] circle (2.5pt) node (00) {} }
        	child {[fill] circle (2.5pt) node (10) {} }
        	child {[fill] circle (2.5pt) node (20) {} }		        	
        	}
        child{ {} 
		child {[fill] circle (2.5pt) node (01) {} }
        	child {[fill] circle (2.5pt) node (11) {} }
        	child {[fill] circle (2.5pt) node(21) {} }        
        	}
        child{[fill] circle (2.5pt) node(2){} }
        ;
     \node [ invisible, below of=00 ]{00};
    \node [ invisible, below of=10 ]{10};
    \node [ invisible, below of=20 ]{20};
    \node [ invisible, below of=01 ]{01};
    \node [ invisible, below of=11 ]{11};
    \node [ invisible, below of=21 ]{21};
    \node [ invisible, below of=2 ]{2};
\end{tikzpicture}
}
\caption{Graphical representation of the context trees $\tau_{quat}$ (left) and $\tau_{ter}$ (right). 
}
\label{fig.tauTQ}
\end{figure}
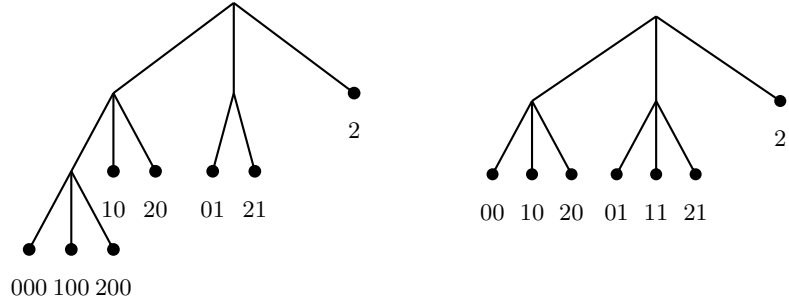

Partitions of the past as described above can be represented by a rooted and labeled \textit{tree} (see Figure \ref{fig.tauTQ}) where each element of the partition is described as a leaf of the tree. 

In the construction described above, for each sequence of past symbols, the algorithm first identifies the corresponding \textit{context} $w$ in the partition $\tau$, where $\tau$ now  represents either $\tau_{ter}$ or $\tau_{qua}$. Once the context $w$ is identified, the algorithm chooses a next symbol $a\in\{0,1,2\}$ using  the transition probability $p(a|w)$. In others terms, each context $w$ in $\tau$ defines a probability measure on $\{0,1,2\}$. The family of transition probabilities indexed by elements of the partition is the second element characterizing the algorithm.

The families of transition probabilities associated to $\tau_{ter}$ and $\tau_{qua}$ are presented in Table \ref{tab:Tran.ProbTQ}.

\begin{table}[h!]
\centering
\begin{subtable}{0.45\textwidth}
%\rowcolors{3}{black!7}{}
\scalebox{0.9}{
\begin{tabular}{c c c c}
\toprule
\multicolumn{4}{c}{\textbf{Quaternary}}\vspace{0.3cm} \\
 \textbf{context} $\mathbf{w}$ & 
 $\mathbf{p(0 | w)}$ & 
 $\mathbf{p(1 | w)}$  &
 $\mathbf{p(2 | w)}$ 
  \\
\toprule
2     &  $\epsilon$ & $1- \epsilon$ & 0\\
21   &  1              & 0                   & 0\\
20   &  1              & 0                   &0 \\
10   &  $\epsilon$ & $1-\epsilon$  & 0\\
01   &   0             & 0                   & 1 \\
200 &  $\epsilon$ & $1-\epsilon$  & 0\\
100 &  0               & 0                  & 1\\
000 &  0              & 0                   & 1\\
\bottomrule
\end{tabular}}
%\caption{Transition probabilities of the Quaternary CTM by context}
\label{table:Simulacao-Valsa}
\end{subtable}
\quad
%\rowcolors{3}{black!10}{} 
\begin{subtable}{0.45\textwidth}
\scalebox{0.9}{
\begin{tabular}{c c c c}
\toprule
\multicolumn{4}{c}{\textbf{Ternary}}\vspace{0.3cm}\\
 \textbf{context} $\mathbf{w}$ & 
 $\mathbf{p(0 | w)}$ & 
 $\mathbf{p(1 | w)}$ & 
 $\mathbf{p(2 | w)}$\\
\toprule
2   &  $\epsilon$ & $1-\epsilon$ & 0\\
21 &  $\epsilon$ & $1-\epsilon$ & 0\\
20 &  $\epsilon$ & $1-\epsilon$ & 0 \\
11 &  0 & 0 & 1\\
10 &  0 & 0 & 1 \\
01 & 0 & 0 & 1\\
00 &  0 & 0 & 1\\
    &      &    &\\
\bottomrule
\end{tabular}}
\end{subtable}
\caption{Transition probabilities of the Quaternary (left) and Ternary (right) chains.}
\label{tab:Tran.ProbTQ}
\end{table}

Using the notion of context tree the neurobiological conjecture can now be rephrased as follows. Is the brain able to identify the context tree
generating the sample of auditory stimuli? From an experimental point of view, the question is
whether it is possible to retrieve each one of the trees presented in Figure
\ref{fig.tauTQ} from the corresponding EEG data.  To deal with this question we
introduce a new statistical model selection procedure described below.

Let $(X_n)$ be either $(X^{ter}_n)$ or $(X^{qua}_n)$ and call $Y_n$ the chunk of EEG data recorded while the volunteer is
exposed to the auditory stimulus $X_n$. Observe that $Y_n$ is a continuous function taking values in $\R^d$, where $d\geq 1$ is the number of electrodes. Its domain is the time interval during which the acoustic stimulus $X_n$ is presented. 
From now on, call $Y^e_n$ the EEG chunk recorded at the electrode $e$ during the exposure to the auditory unit $X_n$.

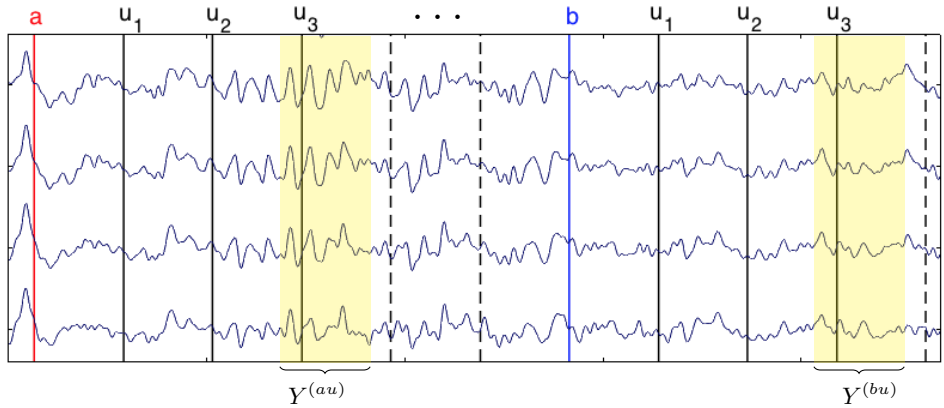
\begin{figure}[h!]
\begin{tikzpicture}
\node (meio) {\includegraphics[scale=0.45]{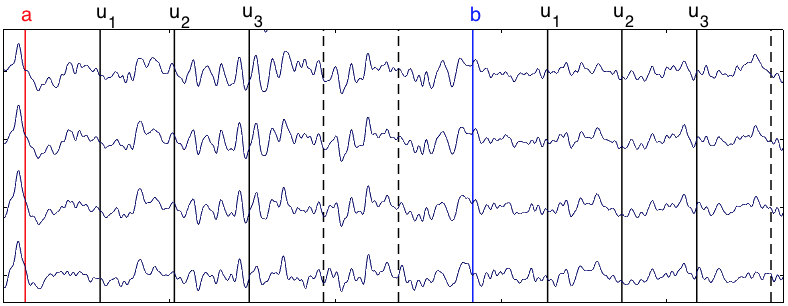}};
\fill [yellow, nearly transparent ](-2.6,-2.4) rectangle (-1.4,1.95);
\fill [yellow, nearly transparent ](4.5,-2.4) rectangle (5.7,1.95);
\draw[snake=brace](-1.4,-2.45) -- (-2.6,-2.45) ;
\draw[snake=brace](5.7,-2.45) -- (4.5,-2.45) ;
\node (aux)[below of=meio, node distance=2.8cm] {};
\node (yk)[left of=aux, node distance=2.1cm] { $Y^{(au)}$};
\node (ykk)[right of=yk, node distance=7.35cm] { $Y^{(bu)}$};
\node (aux2)[above of=aux, node distance=5cm] {};
\node (dots)[left of=aux2, node distance=0.5cm] {{\LARGE  $\ldots$}};
\end{tikzpicture}
\caption{EEG signals recorded from four electrodes. 
The sequence of stimuli is indicated in the top horizontal line. 
Vertical lines indicate the beginning of the successive auditory units. 
The distance between two successive vertical lines is $450$ ms.
Solid vertical lines indicate the successive occurrence times of the string $\mathbf{u}$.
The first yellow strip corresponds to the chunk $Y_n^{(a\mathbf{u})}$ associated to the string $a\mathbf{u}$.
The second yellow strip corresponds to the chunk $Y_n^{(b\mathbf{u})}$ associated to the string $b\mathbf{u}$.
}
\label{fig:ilustra_Yaw}
\end{figure}

The statistical procedure introduced in the paper can be informally described as follows. 
Given a sample $(X_0,Y^e_0), ... , (X_n,Y^e_n)$ of auditory stimuli and associated EEG chunks and for a suitable initial integer $k\geq 1$, do the following.
\begin{enumerate}
\item  For each string $\mathbf{u} = u_1,u_2, ..., u_{k}$, identify all occurrences in the sequence $X_0,X_1, ..., X_n$  of the string $a\mathbf{u}$, obtained by concatenating the symbol $a\in\{0, 1, 2\}$ and the string $\mathbf{u} $. 
\item 
 For each $a\in\{0, 1, 2\}$, define the subsample of all the EEG chunks $Y^e_m=Y_m^{(a\mathbf{u}),e}$ such that $X_{m-k}=a, X_{m-k+1}=u_1, ... ,X_m=u_{k}$ (see Figure \ref{fig:ilustra_Yaw}).
\item  For any pair $a,b\in \{0, 1, 2\}$,
test the null hypothesis that the law of the EEG chunks $Y^{(a\mathbf{u}),e}$ and $Y^{(b\mathbf{u}),e}$ collected at Step 2 are equal.
\begin{enumerate}
\item If the null hypothesis is not rejected for any pair of final symbols $a$ and $b$,  we conclude that the occurrence of $a$ or $b$ before the string $\mathbf{u}$ do not affect the law of EEG chunks.
Then we start again the procedure with the one step shorter sequence $\mathbf{u}= u_{2}, ..., u_{k}$. 
\item If the null hypothesis is rejected for at least one pair of final symbols $a$ and $b$, we conclude that the law of EEG chunks depend on the entire string $a\mathbf{u}$ and we stop the pruning procedure. 
\end{enumerate}
\item 
We keep pruning the sequence $u_1, ..., u_k$ until the null-hypothesis is reject for the first time.
\item
Call $\hat \tau^e_n$ the tree constituted by the strings which remained after the pruning procedure.
\end{enumerate}
The question is whether $\hat \tau^e_n$ coincides with the context tree $\tau$ generating the sequence of auditory stimuli. 

An important technical issue must be clarified at this point, namely, how to test the equality of the laws of two subsamples of EEG chunks. 
This is done using the projective method informally explained below.

Suppose we have two samples of random functions, each sample composed by independent realizations of some fixed law. In order to test whether the two samples are generated by the same law,  we choose at random a ``direction'' and project each  function in the samples in this direction. This produces  two new samples of real numbers. Now we test whether the samples of the projected real numbers have the same law. 
Under suitable conditions, a theorem by \cite{Cuesta2006} ensures that for almost all directions if the test does not reject the null hypothesis that the projected samples have the same law, then the original samples also have the same law.

The arguments informally sketched in this section are formally developed in Sections \ref{sec.HCTMselection} and \ref{sec:case_study} and in the Appendix.

\section{Stochastic processes driven by context tree models}
\label{sec:HCTM}

Let $A$  be a finite alphabet. Given two integers $m,n\in \Z$ with $m\leq n$, the string $(u_m,\ldots, u_n)$ of symbols in $A$ is often denoted by $u^n_m$; its length is $\ell(u^n_m)=n-m+1$. The empty string is denoted by $\emptyset$ and its length is $\ell(\emptyset)=0$. Fixed two strings $u$ and $v$ of elements of $A$, we denote by $uv$ the string in $A^{\ell(u)+\ell(v)}$ obtained by the concatenation of $u$ and $v$. By definition $u\emptyset=\emptyset u=u$ for any string $u\in A^{\ell(u)}$. 
The string $u$ is said to be a \textit{suffix}\index{suffix} of $v$ if there exists a string $s$ satisfying $v=su$. This relation will be denoted by $u\preceq v$. When $v\neq u$ we say that $u$ is a \emph{proper suffix} of $v$ and write $u\prec v$.
Hereafter, the set of all  finite strings of symbols in $A$ is denoted by $A^*:=\bigcup_{k=1}^{\infty}A^k$.  
For any finite string
$w=w^{-1}_{-k}$ with $k\geq 2$, we write
$\mbox{suf}(w)$  to denote the one-step shorter string $w^{-1}_{-k+1}$. 
\begin{definition}
\label{def.contexttree}
A finite subset $\tau$ of $A^*$ is a \textit{context tree} if it satisfies the following conditions:
\begin{enumerate}
\item \textit{Suffix Property.} For no $w\in \tau$ we have $u\in \tau$ with $u\prec w$ .
\item \textit{Irreducibility.} No string belonging to $\tau$ can be replaced by a proper suffix without violating the suffix property.
\end{enumerate}
\end{definition}

The set $\tau$ can be identified with the set of leaves of a rooted tree with a finite set of labeled branches. The elements of $\tau$ will be always denoted by $w,u,v, \ldots$. A context tree is said to be complete if each node but the leaves has $|A|$ children. 
\begin{remark}
Context trees are not necessarily complete. For instance the tree $\tau_{qua}$ corresponding to the Quaternary chain is not complete, see Figure \ref{fig.tauTQ}. This is due to the fact that by construction the string $11$ can not appear in that chain. The important fact about context trees is that they define a partition of the set of all possibles sequences of past symbols. This is assured by the irreducibility condition. 
\end{remark}

The height of the context tree $\tau$ is defined as 
$
\ell(\tau)=\max\{\ell(w) \,:\, w \in \tau\}.
$
In the present paper we only consider context trees with finite height.

\begin{definition}
\label{def.ordercontextree}
Let $\tau$ and $\tau'$ be two context trees. We say that 
$\tau$ is smaller than $\tau'$ and write
$\tau \preceq \tau\p$, if for every $w\p\in\tau\p$ there exists $w\in\tau$ such that $w \preceq w\p$. 
\end{definition}

Given a context tree $\tau$, let $p=\{p(\cdot\mid w): w\in\tau\}$ be a family of probability measures on $A$ indexed by the elements of $\tau$. 
\begin{definition}
The pair $(\tau,p)$ will be called a \textit{probabilistic context tree} on $A$. Each element of $\tau$ will be called a \textit{context}.
\end{definition}

\begin{definition}
\label{def:irreducible}
A probabilistic context tree $(\tau,p)$ with height $\ell(\tau)=k$ is {\it irreducible} if for any $a^{-1}_{-k}\in A^k$ and $b\in A$ there exist a positive integer $n=n(a^{-1}_{-k},b)$ and symbols $a_0, a_1, \ldots, a_n=b\in A$ such that 
$$
p(a_0|c_{\tau}(a^{-1}_{-k}))>0, p(a_1|c_{\tau}(a_0a^{-1}_{-k}))>0, \ldots, p(a_n|c_{\tau}(a_{n-1},\ldots, a_0a^{-1}_{-k}))>0. 
$$ 
\end{definition}

\begin{definition}
\label{def-CMAV}
Let $(\tau,p)$ be a probabilistic context tree on $A$. A stochastic chain $(X_n)_{n \in \N}$ taking values in $A$ is called a \textit{context tree model} compatible with $(\tau,p)$ if 
\begin{enumerate}
\item \label{def.CTM} for any $n\geq \ell(\tau)$ and any finite string $x^{-1}_{-n}\in A^{n}$ such that $P\big(X^{n-1}_{0}=x^{-1}_{-n}\big)>0$, it holds that
\begin{equation}
\label{def:scmvl}
P\Big( X_n=a\mid X^{n-1}_{0}=x^{-1}_{-n}\Big) =p\big(a\mid c_{\tau}\big(x^{-1}_{-n}\big)\big) \ \  \mbox{for all} \ a\in A,
\end{equation}
where $c_{\tau}\big(x^{-1}_{-n}\big)$ is the only context in $\tau$ which is a suffix of $x^{-1}_{-n}$.
\item For any $1\leq j < \ell(c_{\tau}\big(x^{-1}_{-n}\big))$, there exists $a\in A$ such that
$$
P\Big( X_n=a\mid X^{n-1}_{0}=x^{-1}_{-n}\Big)\neq P\Big( X_n=a\mid X^{n-1}_{n-j}=x^{-1}_{-j}\Big).
$$
\end{enumerate}
\end{definition}

Section \ref{sec:PresentationOfOurApproach} presents the two context tree models which were used in our experimental protocol, namely  the chains $(X^{ter}_n)$ and $(X^{qua}_n)$ having as context trees $\tau_{ter}$ and $\tau_{qua}$ described in Figure \ref{fig.tauTQ} and corresponding transition probabilities described in Table \ref{tab:Tran.ProbTQ}.

\begin{definition} 
\label{def.HCTM}
Let $A$ be a finite alphabet, $(\tau,p)$ a probabilistic context tree on $A$,  $(F,\mathcal{F})$ a measurable space and $(Q^w: w\in \tau)$ a family of probability measures on  $(F,\mathcal{F})$. The bivariate stochastic chain $(X_n, Y_n)_{n \in \N}$ taking values in $A\times F$ is a \textit{stochastic process driven by a context tree model} compatible with $(\tau,p)$  and $(Q^{w}: w \in \tau)$ if the following conditions are satisfied,
\begin{enumerate}
\item $(X_n)_{n \in \N}$ is a context tree model compatible with $(\tau, p)$.
\item For any integers $\ell(\tau)\leq m\leq n$, any string $x_{m-\ell(\tau)+1}^n\in A^{n-m+\ell(\tau)}$ and any sequence $J_m, \ldots, J_n$ of $\mathcal{F}$-measurable sets,
{\small
$$
P\big( Y_{m} \in J_m , \ldots , Y_n \in J_n| X^n_{m-\ell(\tau)+1}=x_{m-\ell(\tau)+1}^n\big)\!=\!\!\prod_{k=m}^n Q^{c_{\tau}(x_{k-\ell(\tau)+1}^{k})}(J_k),
$$}
where $c_{\tau}(x_{k-\ell(\tau)+1}^{k})$ is the context in $\tau$ assigned to the string of symbols $x_{k-\ell(\tau)+1}^{k}$.
\end{enumerate}
\end{definition}

\begin{definition}
A stochastic process driven by a context tree mode compatible with $(\tau,p)$  and $(Q^{w}: w \in \tau)$ is {\it identifiable} if for any pair of contexts $w\in\tau$ and $u \in\tau$ such that suf$(w)$=suf$(u)$, we have $
Q^w\neq Q^{u}.
$
\end{definition}

The process  $(X_n)$ will be called the  {\it stimulus chain} and $(Y_n)$ will be called the {\it response chain.}

Using the notion of stochastic chain driven by a context tree model our experimental protocol can be now formally described as follows. 
\begin{itemize}
\item The  stimulus chain $(X_n)$ is either the Ternary or the Quaternary chain producing the sequence of the auditory stimuli defined in Section \ref{sec:PresentationOfOurApproach}. 
\item Each element $Y_n$ of the the response chain $(Y_n)$ is the EEG chunk  recorded while the volunteer is exposed to auditory stimulus $X_n$. Thus $Y_n=(Y_n(t),t\in [0,T])$ is a real function, where $T$ is the time distance between the onsets of two consecutive auditory stimuli. The sample space $F$ is the Hilbert space $L^2([0,T])$ of  real-valued square integrable functions over $[0,T]$ and $\mathcal{F}$ is the Borel $\sigma$-algebra on $F$. 
\item Finally, $(Q^w, w\in \tau)$ is a family of probability measures on $L^2([0,T])$ describing the laws of the EEG chunks.
\end{itemize}

\section{Statistical selection for stochastic processes driven by context tree models}
\label{sec.HCTMselection}
Let $F=L^2([0,T])$ be the Hilbert space of real-valued square integrable functions over $[0,T]$ for some $T>0$ and $\mathcal{F}$ be the Borel $\sigma-$algebra on $L^2([0,T])$. Moreover, let $(\bar{\tau},\bar{p})$ be a probabilistic context tree on a finite alphabet $A$ and $(\bar{Q}^w:w\in\bar{\tau}) $ be a family of probability measures on $(F,\mathcal{F})$.
 Finally, let  
$(X_0,Y_0),\ldots,(X_{n},Y_{n})$, with $X_k\in A$ and $Y_k\in F$ for $0\leq k\leq n$, be a sample produced by a stochastic process driven by a context tree model compatible with  $(\bar{\tau},\bar{p})$ and  $(\bar{Q}^w:w\in\bar{\tau}) $.   
Before presenting our statistical selection procedure we need two more definitions.
\begin{definition}
\label{def:branch}
Let $\tau$ be a context tree and fix a finite string $s\in A^*$. We define the \textit{branch} in $\tau$ induced by $s$ as the set $B_{\tau}(s)=\{w\in\tau : w\succ s\}$. The set $B_{\tau}(s)$ is called a \textit{terminal branch} if for all $w\in B_{\tau}(s)$ it holds that $w=as$ for some $a\in A$.
\end{definition}
Given a sample  $X_0,\ldots, X_{n}$ of symbols in $A$ and a finite string $u\in A^*$, the number of occurrences of $u$ in the sample $X_0,\ldots, X_{n}$ is defined as 
$$
N_n(u)=\sum_{m=l(u)-1}^{n}1\{X^{m}_{m-\ell(u)+1}=u\}.
$$
\begin{definition}
\label{def.:admissible}
Given integers $n>L\geq 1$, an \textit{admissible context tree} of maximal height L for the sample $X_0,\ldots,X_{n}$ of symbols in $A$, is any context tree $\tau$ satisfying
\begin{enumerate}
\item $w\in\tau$ if and only if $\ell(w)\leq L$ and $N_n(w)\geq 1$.
\item Any string $u\in A^*$ with $N_n(u)\geq 1$ is  a suffix of some $w\in\tau$ or has a suffix $w\in\tau$.
\end{enumerate} 
\end{definition}

For any pair of integers $1 \leq L <n$ and any string $u\in A^*$ with $\ell(u)\leq L$,
call $I_n(u)$ the set of indexes belonging to $\{\ell(u)-1,\ldots, n\}$ in which the string $u$ appears in sample $X_0,\ldots, X_{n}$, that is
$$
I_n(u)=\{\ell(u)-1\leq m\leq n: X^{m}_{m-\ell(u)+1}=u\}.
$$
Observe that by definition $|I_n(u)|=N_n(u)$. If $I_n(u)=\{m_1,\ldots, m_{N_n(u)}\}$, we set $Y^{(u)}_k=Y_{m_{k}}$ 
for each $1\leq k\leq N_n(u)$. Thus, $Y^{(u)}_1,\ldots,Y^{(u)}_{N_n(u)}$ is the {\it subsample of} $Y_0, \ldots, Y_{n}$ \textit{induced by the string} $u$.

Given $u\in A^*$ such that $N_n(u)\geq 1$ and $h\in L^2([0,T])$, we define the empirical distribution associated to the projection of the sample $Y^{(u)}_1,\ldots,Y^{(u)}_{N_n(u)}$ onto the direction $h$ as
$$
\hat{Q}^{u,h}_n(t)=\frac{1}{N_n(u)}\sum_{m=1}^{N_n(u)}1_{(-\infty,t]}(\langle Y^{(u)}_m, h\rangle), \ t \in \R, 
$$ 
where for any pair of functions $f, h \in L^2([0,T])$, 
$$
\langle f, h\rangle=\int_{0}^Tf(t)h(t)dt.
$$
For a given pair $u,v\in A^*$ with $\max\{\ell(u),\ell(v)\}\leq L$ and $h\in L^2([0,T])$, the Kolmogorov-Smirnov distance between the empirical distributions $\hat{Q}^{u,h}_n$ and $\hat{Q}^{v,h}_n$ is defined by
$$
KS(\hat{Q}^{u,h}_n,\hat{Q}^{v,h}_n)=\sup_{t\in \R}|\hat{Q}^{u,h}_n(t)-\hat{Q}^{v,h}_n(t)|.
$$
Finally, we define for any pair $u,v\in A^*$ such that $\max\{\ell(u),\ell(v)\}\leq L$ and $h\in L^2([0,T]),$
{\small
$$
D^h_n((Y_1^{(u)},\ldots, Y_{N_n(u)}^{(u)}),(Y_1^{(v)},\ldots, Y_{N_n(v)}^{(v)}))=\sqrt{\frac{N_n(u)N_n(v)}{N_n(u)+N_n(v)}}KS(\hat{Q}^{u,h}_n,\hat{Q}^{v,h}_n).
$$
}
Our selection procedure can be now described as follows. Fix an integer $1\leq L<n$ and 
let $\mathcal{T}_n$ be the largest admissible context tree of maximal height $L$ for the sample $X_0,\ldots, X_{n}$. The largest means that  if $\tau$ is any other admissible context tree of maximal height $L$ for the sample $X_1^n $, then $\tau\preceq \mathcal{T}_n.$

For any string $u\in A^*$ such that ${B}_{\mathcal{T}_n}(u)$ is a terminal branch
we test the null hypothesis
\begin{equation}
\label{testH_0}
H^{(u)}_0\!\!: \mathcal{L}\big(Y^{(au)}_1,\ldots,Y^{(au)}_{N_n(au)}\big)
\!=\!\mathcal{L}\big(Y^{(bu)}_1,\ldots,Y^{(bu)}_{N_n(bu)}\big),  \, \forall \,  au,bu\in {B}_{\mathcal{T}_n}(u)
\end{equation}
using the test statistic 
\begin{equation}
\label{prun.funEEG}
\Delta_n(u)\!=\!\Delta^W_n(u)\!=\!\max_{a,b\in A}D^W_n\big((Y_1^{(au)},\ldots, Y_{N_n(au)}^{(au)}),(Y_1^{(bu)},\ldots, Y_{N_n(bu)}^{(bu)})\big),
\end{equation} 
where $W=(W(t):t\in[0,t])$ is a realization of a Brownian motion in $[0,T]$. We reject the null hypothesis $H_0^{(u)}$ when $\Delta_n(u)>c$, where $c>0$ will be specified later. When the null hypothesis $H_0^{(u)}$ is not rejected,  we prune the branch $B_{\mathcal{T}_n}(u)$ in $\mathcal{T}_n$ and set as a new candidate context tree
$$
\mathcal{T}_n=\big(\mathcal{T}_n\setminus {B}_{\mathcal{T}_n}(u)\big) \cup \{u\}.
$$
On the other hand, if the  null hypothesis $H_0^{(u)}$ is rejected, we keep $B_{\mathcal{T}_n}(u)$ in $\mathcal{T}_n$ and stop testing $H_0^{(s)}$ for strings $s\in A^*$ such that $s\preceq u$. 

In each pruning step, take a string $s\in A^*$ which induces a terminal branch in $\mathcal{T}_n$ and has not been tested yet.
This pruning procedure is repeated until no more pruning is performed. 
We denote by $\hat{\tau}_n$ the final context tree obtained by this procedure. 
The formal description of the above pruning procedure is provided in Algorithm 1 as a pseudo code.

\begin{algorithm}[h!]
\renewcommand{\algorithmicrequire}{\textbf{Input:}}
\renewcommand{\algorithmicensure}{\textbf{Output:}}
\caption{}
\label{alg:pruning}
\begin{algorithmic}[1]
\REQUIRE A sample $(X_0,Y_0),\ldots,(X_{n},Y_{n})$ with $X_k\in A$ and $Y_k \in F$ for $0\leq k\leq n$, a positive threshold $c$ and a positive integer $L$.
\ENSURE A tree $\hat{\tau}_n$
\State $\tau \leftarrow \mathcal{T}_n$ 
\State Flag$(s)\leftarrow $ ``not visited'' for all string $s$ such that $s\preceq w\in \mathcal{T}_n$
\FOR{k in L to 1 }
\WHILE{$\exists s \in \tau$: $\ell(s)=k$,  Flag$(s) = $ ``not visited'' and $B_{\tau}(s)$ is a terminal branch}
\State Choose a $s$ such that $\ell(s)=k$,  Flag$(s) = $ ``not visited'' and $B_{\tau}(s)$ is a terminal branch
\State Compute the test statistic $\Delta_n(s)$ to test $H^{(s)}_0$ 
\IF{$\Delta_n(s)>c$}
\State Flag$(u) \leftarrow$ ``visited'' $\forall u\preceq s$ 
\ELSE{}
\State $\tau\leftarrow (\tau\setminus B_{\tau}(s))\cup \{s\}$
\ENDIF
\ENDWHILE
\ENDFOR
\State{\textbf{Return} $\hat{\tau}_n=\tau.$}
\end{algorithmic}
\end{algorithm}

To state the consistency theorem  we need the following definitions.

\begin{definition}
\label{ass:1}
A probability measure $P$ defined on $(L^{2}([0,T]), \mathcal{B}(L^{2}([0,T])))$ satisfies {\em Carleman condition} if all the absolute moments $m_k=\int ||h||^kP(dh)$, $k\geq 1,$ are finite and
$$
\sum_{k\geq 1}m_k^{-1/k}=+\infty.
$$
\end{definition}

\begin{definition}
Let $P$ be a  probability measure on $(L^{2}([0,T]), \mathcal{B}(L^{2}([0,T])))$. 
We say that $P$ is {\em continuous} if $P^h$ is continuous for any $h\in L^{2}([0,T])$, where
$P^h$ is defined  by 
$$P^h((-\infty, t])=P(x\in L^{2}([0,T]): \langle x, h \rangle\leq t), \ t\in\R.$$

Let $V$ be a finite set of indexes and $(P_i:i\in V)$ be a family of probability measures on  $(L^{2}([0,T]), \mathcal{B}(L^{2}([0,T])))$. We say that $(P_i:i\in V)$ is continuous if for all $i\in V$, the probability measure $P_i$ is continuous.
\end{definition}

\begin{thm}
\label{thm:2}
Let  $(X_0,Y_0),\ldots, (X_{n},Y_{n})$ be a sample produced by a identifiable stochastic process driven by a context tree model compatible with $(\bar{\tau},\bar{p})$ and $(\bar{Q}_w:w\in\bar{\tau})$, and let  $\hat{\tau}_n$ be the context tree selected from the sample by Algorithm \ref{alg:pruning} with  $L\geq \ell(\bar{\tau})$. If $(\bar{\tau},\bar{p})$ is irreducible and $(\bar{Q}_w:w\in\bar{\tau})$ is continuous and satisfies Carleman condition, then for any $\epsilon>0$ there exists a threshold $c>0$ such that for all $n$ large enough  
$$
P(\hat{\tau}_{n}\neq \bar{\tau})\leq \epsilon.
$$
\end{thm}
The proof of Theorem \ref{thm:2} is postponed to Appendix \ref{App:1}.

\section{Case study: retrieving context trees from EEG data}
 \label{sec:case_study}
In this section we apply our statistical procedure to an original EEG dataset collected by the authors using our new experimental protocol. We also use the EEG data as input for a simulation study.

A total of 20 healthy volunteers (9 female, mean age 30 y., standard deviation 6.8 y., 18 right handed) was evaluated. None of them have reported any neurological pathology. The volunteers signed an informed consent term, after the nature of the study had been explained. This experimental protocol was approved by the local ethics committee ({\it Plataforma Brasil} process number 22047613.2.0000.5261).

\subsection{Experimental protocol}
\label{subsec:protocol}
The experiment consisted in exposing volunteers to sequences of auditory stimuli defined as strong beats, weak beats and silent units, indicated respectively by symbols $2, 1$ and $0$. The sequence of auditory stimuli were produced by the Ternary and Quaternary chains defined in Section \ref{sec:PresentationOfOurApproach}. In the experiment, the parameter  $\epsilon$ appearing in the transition probabilities in Table \ref{tab:Tran.ProbTQ} took the value $\epsilon=0.2$.  

Besides the Ternary and Quaternary chains we also used sequences of independent auditory stimuli taking the values $0,1$ and $2$ with probability $1/3$. The goal of introducing this i.i.d. sequence of stimuli was to {\it shuffle cards} before the volunteer is exposed to a next sample. 

The volunteer was exposed to two 12 min blocks of samples generated by each one of the three stochastic chains. The blocks were separated from each other by a period of time ranging from 5 to 10 min, during which data collection was interrupted.  
Each block was a concatenation of three 1 min sequences of auditory units generated independently by the same stochastic chain. Each sequence of auditory units was separated from the next one by a 15 seconds silent interval.  

All volunteers were exposed to two different orderings, either Ternary, Independent, Quaternary
or Quaternary, Independent, Ternary.
For half of the volunteers the starting block was Ternary, Independent, Quaternary and the second block was Quaternary, Independent and Ternary. The inverse ordering was used with the other half, to balance possible order effects. 

Presentation software  (Presentation Mixer as a Primary Buffer and a Sound card: SoundMAX HD Audio) was used to play the auditory sequences  through a headset. The loudness of the stimuli was individually regulated before the experiment start by presenting the strong beat and asking the volunteer to adjust it up to a comfortable level (Range: 0.1-0.3 dB). 

\subsection{Data acquisition and pre-processing}
\label{subsec:dataacquistion}
EEG recording was performed  by means of a 128 channels system (Geodesic HidroCel GSN 128 EGI, Electrical Geodesic Inc.). The electrode cap, previously immersed in saline solution (KCl), was  dressed into the volunteer's scalp. Volunteers  were instructed to close their eyes and remain quiet throughout the experiment.

The EEG signal was amplified with a nominal gain of 20 times. The acquisition was performed in a sampling frequency of 250 Hz. During acquisition  the signal was  analogically filtered (Butterworth first order band-pass filter of 0.1-200Hz; Geodesic EEG System 300, Electrical Geodesic Inc.). The electrode positioned on the vertex (Cz) was used as reference.   

 The data  was preprocessed offline using  EEGLAB (\cite{Delorme-eeglab:04}) running in MATLAB environment (MathWorks, Natick, MA, version R2012a). Signals were filtered with a Butterworth fourth order band-pass  filter  of $1$-$30$ Hz. Artifacts above and below 100 $\mu$V were removed. The  data was  then segmented into events of $450$ ms, each one indexed by the  corresponding  auditory unit. Finally, baseline correction was performed using the signal collected 50 ms before each event starts.

\subsection{Data analysis}
\label{sub.sec:datanalysis}
The statistical analysis was perfomed using EEG data recorded at the following scalp positions according to the 
$10$-$20$
system: FP2 , FP1, F7, F3, Fz, F4, F8, T3, C3, C4, T4, T5, P3, Pz, P4, T6, O1,  O2. This set of electrodes will be denoted $\mathcal{E}=\{e_1,e_2,\ldots, e_{18}\}$. We will write $\mathcal{V}=\{v_1,v_2,\ldots,v_{20}\}$ to denote the set of 20 volunteers. Finally, $\mathcal{T}=$\{$\tau_{ter}$,\ $\tau_{qua}$\} is the set of context tree models used in the experiment.  

The sequence of auditory units produced by the context tree model $\tau\in\mathcal{T}$ and presented to  volunteer $v\in\mathcal{V}$ is denoted by $X^{\tau,v}_0,\ldots,X^{\tau,v}_{n}$.  In the experiment $n=799$, the length of the time interval between consecutive auditory units is $450 \ ms$ and the total time of exposure to each context tree model is 6 min. 
The sequence of EEG chunks recorded at electrode $e\in \mathcal{E}$ is denoted by $Y^{\tau,v,e}_0,\ldots,Y^{\tau,v,e}_{n}.$ Recall that $Y^{c,v,e}_k:[0,T]\to \R$ is the EEG chunk recorded while the k-th stimulus unit $X^{\tau,v}_k$ occurs. Since the sampling frequency of the EEG acquisition is $250 \ Hz$ and $T=450 \ ms$, any EEG chunk is a real vector with length $113$. 

Therefore, the sample associated to the context tree model 
$\tau\in\mathcal{T}$, volunteer $v\in\mathcal{V}$ and electrode $e\in \mathcal{E}$ is
\begin{equation}
\label{dataset}
(X^{\tau,v}_0, Y^{\tau,v,e}_{0})\ldots, (X^{\tau,v}_{n},Y^{\tau,v,e}_{n}),
\end{equation}
where $n=799$ and $Y^{\tau,v,e}_k\in \R^{113}$ for all $0\leq k\leq n$.

To make the test more stable, instead of using only one random projection we will take several random projections in Algorithm \ref{alg:pruning}( see \cite{Cuesta2006} and \cite{Cuestaetal:07}). 
To do this we replace step 6 of Algorithm \ref{alg:pruning} by
\begin{enumerate}
\item[$6$-i.] generate $N$ independent realizations $W_1,\ldots, W_N$ of a Brownian motion $W=(W(t):t\in[0,T])$. 
\item[$6$-ii.]  compute the test statistics $\Delta^{W_1}_n(s),\ldots, \Delta^{W_N}_n(s).$
\item[$6$-iii.]  define the statistic  
$$
\tilde{\Delta}_n(s)=\sum_{m=1}^N 1\{\Delta^{W_m}_n(s)>c\}.
$$
\end{enumerate}
and step 7 by
\begin{enumerate}
\item[$7.$] {\bf if} $\tilde{\Delta}_n(s)>C$, {\bf then} perform step 8 of Algorithm \ref{alg:pruning}. {\bf Else}, perform steps 10 and 11 of Algorithm \ref{alg:pruning}. 
\end{enumerate}

The resulting algorithm will be called Algorithm 2.  
We call $\hat{\tau}^{\tau,v,e}_n$ the context tree selected by Algorithm 2 from the sample $(X^{\tau,v}_0, Y^{\tau,v,e}_0)\ldots, (X^{\tau,v}_{n},Y^{\tau,v,e}_{n})$.

The choice of constants $c$ and $C$ is based on Proposition \ref{prop:bootstrap} below.  

\begin{proposition}
\label{prop:bootstrap}
For any string $s\in A^*$ and integer $N\geq 1$, consider the random variables   
$\Delta^{W_1}_n(s),\ldots, \Delta^{W_N}_n(s)$ defined in \eqref{prun.funEEG} where $W_1, \ldots W_N$ are independent realizations of a Brownian motion $W=(W(t): t\in [0,T])$. For any $\alpha\in (0,1)$, we set  $c_{\alpha}=(-1/2\ln(\alpha/2))^{1/2}$. Under the null assumption  $H_0^{(s)}$ defined in \eqref{testH_0}, it follows that  as $n\to\infty$,
$$
\sum_{m=1}^N 1\{\Delta^{W_m}_n(s)>c_{\alpha}\} \xrightarrow{d} Bin(N,\alpha),
$$
where $\xrightarrow{d}$ means convergence in distribution and $Bin(N,\alpha)$ denotes a random variable with Binomial distribution of parameters $N$ and $\alpha.$
\end{proposition}
The proof of Proposition \ref{prop:bootstrap} is postponed to Appendix \ref{App:3}.

\subsection{Simulation study}
\label{sec:AlgorArvore-fw}
For each $v\in\mathcal{V}$ and $e\in \mathcal{E}$, we want to simulate $(\tilde{X}^{v}_n, \tilde{Y}^{v,e}_n)$ corresponding to the stimulus and response chains respectively for either the Ternary or Quaternary case.  From now on, in this description, $\bar{\tau}$ denotes either $\tau_{ter}$ or $\tau_{qua}.$  
The simulation is done as follows.
\begin{itemize}
\item For each $v\in\mathcal{V}$, the stimuli chain $(\tilde{X}^v_n)$ is generated by $\bar{\tau}$ and its corresponding transition probabilities. We do this independently for each $v\in\mathcal{V}.$
\item To simulate the response chain $(\tilde{Y}^{v,e}_n)$ we use the family of distributions $(\tilde{Q}^{w,e,v})_{w\in\bar{\tau}}$ defined as follows.
\begin{itemize}
\item For each $v\in\mathcal{V}$, $e\in \mathcal{E}$ and  $w\in\bar{\tau}$, we collect together all the EEG chunks recorded from volunteer $v$ at electrode $e$ at each realization of context $w$. This means that we consider the EEG chunk recorded during the exposure of the volunteer to the last unit of $w$, concluding a complete realization of $w$. 
\item  Let $F^{w,e,v}$ be the set of these EEG chunks. The distribution $\tilde{Q}^{w,e,v}$ is defined as the uniform distribution on $F^{w,e,v}$. 
\end{itemize}
\item For each  $\ell(\bar{\tau})\leq m$, we choose the value of
$\tilde{Y}^{e,v}_m$ according to $\tilde{Q}^{w,e,v}$  where $w=c_{\tau}(\tilde{X}^v_{0}, \ldots, \tilde{X}^v_{m} )$.
\end{itemize}

Algorithm 2 applied to the  sample $(\tilde{X}^{v}_0,\tilde{Y}^{v,e}_0), \ldots, (\tilde{X}^{v}_n, \tilde{Y}^{v,e}_n)$ selects a context tree for volunteer $v\in\mathcal{V}$ and electrode $e\in \mathcal{E}$. 

In the simulation study, we used the parameters $n=799$, $L=3$, $N=100$, $c_{\alpha}=(-1/2\ln(\alpha/2))^{1/2}$ with $\alpha=0.05$, and $C=9$ for both stochastic chains.  The value
$C=9$ corresponds to $0.95$-quantile of a Binomial distribution of parameters $N=100$ and $\alpha=0.05$. The results of the simulation study are reported in Table \ref{tab:simulacao}.

\begin{table}[h!]
\includegraphics[scale=.4]{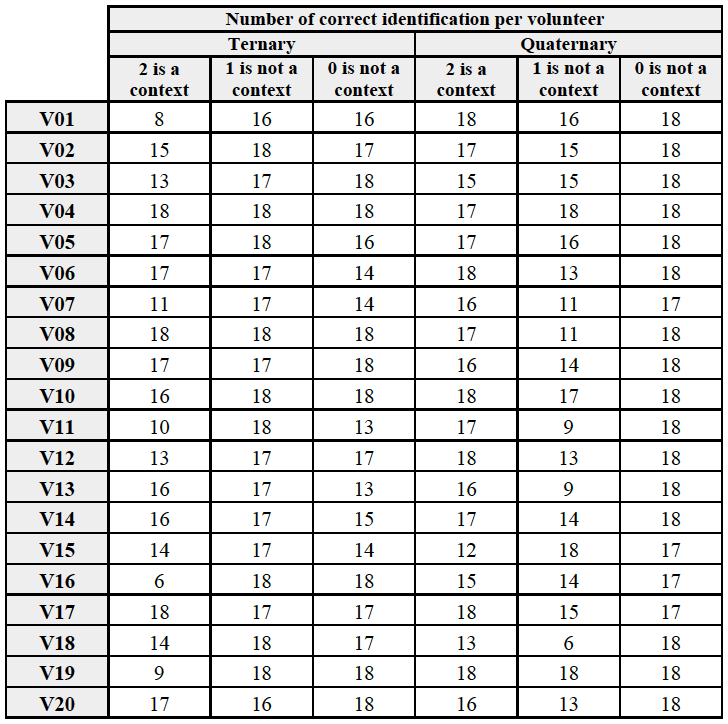} 
\caption{
For both Ternary and Quaternary chains all the 18 electrodes selected a non-empty context tree.
The results in the first three columns correspond to the Ternary chain. The first one indicates the number of electrodes that correctly identify $2$ as a context.  The second and third columns indicate the number of electrodes that correctly identify that $1$ or $0$ are not contexts. The results in the last three columns correspond to the Quaternary chain. 
}
\label{tab:simulacao} 
\end{table}

\subsection{Retrieving context trees from experimental data}
The results obtained from the analysis of the EEG signals recorded while the volunteers were exposed to the Quaternary and Ternary chains are summarized in Table \ref{Table:results}. In the analysis, we used the same parameters $L=3$, $N=100$, $c_{\alpha}=(-1/2\ln(\alpha/2))^{1/2}$ with $\alpha=0.05$, and $C=9$ for both stochastic chains. 

\begin{table}[h!]
\centering
\includegraphics[scale=0.5]{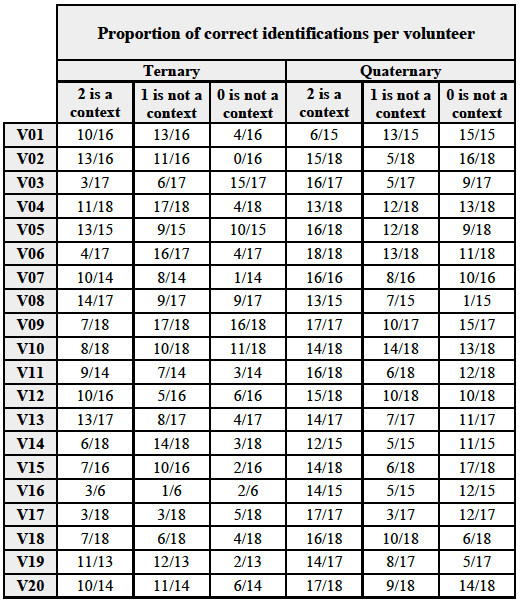}
\caption{
The results in the first three columns correspond to the Ternary chain. The first one indicates the proportion of electrodes that correctly identify $2$ as a context.  The second and third columns indicate the proportion of electrodes the correctly identify that $1$ or $0$ are not contexts. The results in the last three columns correspond to the Quaternary chain. 
All the proportions refer to the number of electrodes that select a non-empty context tree.}
\label{Table:results}
\end{table}

The results in Table \ref{Table:results} show for instance that for volunteer V01 in the Ternary chain, from $18$ selected trees, $16$ are non-empty. Among the non-empty trees, $10$ correctly have $2$ as a context, while in $13$ of them $1$ correctly is not a context. In the Quaternary chain, for volunteer V06, for instance, all the $18$ electrodes selected non-empty trees and all of them correctly have $2$ as a context, while $13$ of them (respectively $11$ of them) correctly do not have $1$ (respectively $0$)  as a context. 

We only reported the results of comparisons between EEG chunks corresponding to strings with length not larger than 2. Actually, the results of the statistical analysis done with  EEG chunks corresponding to larger strings of auditory units 
are less satisfactory. This is due to the small size of the samples used in each {\sl prune-or-keep} decision.  For instance in the case of the Quaternary chain, 14 volunteers failed to identify $000$ as a context and 7 volunteers incorrectly decide not to prune the branch $020$. This is not surprising when we observe that the strings $000$ and $020$ appear at most 14 and 15 times respectively in the sequence of stimuli presented to the volunteers. 
 In the Ternary chain, the strings $020$ and $002$ appear at most 17 and 15 times respectively.

In general, the results obtained in the Ternary case are less impressive than those obtained for the Quaternary case. This could be a consequence of the fact that in the Ternary case the two random units (which could take the values 0 or 1) occur successively, while in the Quaternary case the two random units are separated by deterministic units.

Interestingly, a clear intersubject variability concerning both the number of electrodes performing the expected context detection and the number of electrodes which select a non-empty tree was identified. The reasons for such variability deserve further future exploration.  

The simulation results were qualitatively similar to those obtained with empirical data, albeit with a somewhat higher performance, as expected.  In conclusion, the new functional data statistical selection procedure finds results which support the conjecture that the brain effectively identifies the context tree generating the sample of stimuli.

The EEG data and the codes of the algorithms used both in the statistical analyses 
and simulation study can be downloaded from 
\url{https://goo.gl/tfBjxh} and  \url{https://github.com/neuromat/eeg-tree-algorithms}  respectively.

\appendix

\section{Mathematical proofs}
\label{sec:Append}

\subsection{Proof of Theorem \ref{thm:2}}
\label{App:1}

The proof of Theorem \ref{thm:2} will be an direct consequence of Propositions \ref{prop:1} and \ref{prop:2}  presented below. 
\begin{proposition}
\label{prop:1}
Let $(\bar{\tau},\bar{p})$ be a  probabilistic context tree and $\{\bar{Q}^w:w\in\bar{\tau}\}$ be a family of probability distributions on $(L^{2}([0,T]), \mathcal{B}(L^{2}([0,T])))$ satisfying the assumptions of Theorem \ref{thm:2}. Let  $\alpha\in (0,1)$ and set $c_{\alpha}=(-1/2\ln(\alpha/2))^{1/2}$. For any integer $L\geq \ell(\bar{\tau})$, context $w\in\bar{\tau}$,
direction $h\in F\setminus\{0\}$, and
 strings $u, v\in \cup_{k=1}^{L-\ell(w)}A^{k}$ such that $w\preceq u$ and $w\preceq v$,  it holds that
$$
\lim_{n\to \infty}P(D_n^h((Y_1^{(u)},\ldots, Y_{N_n(u)}^{(u)}),(Y_1^{(v)},\ldots, Y_{N_n(v)}^{(v)}))>c_{\alpha})=\alpha.
$$
\end{proposition}
\begin{proof}
Since both $N_n(u)$ and $N_n(v)$ tend $P$-a.s. to $+\infty$ as $n$ diverges,
Theorem  3.1(a) of \cite{Cuesta2006} implies that the distribution of $D_n^h((Y_1^{(u)},\ldots, Y_{N_n(u)}^{(u)}),(Y_1^{(v)},\ldots, Y_{N_n(v)}^{(v)}))$ is independent of the strings $u$ and $v$, and also of the direction $h\in F\setminus\{0\}$. 
It also implies that $D_n^h((Y_1^{(u)},\ldots, Y_{N_n(u)}^{(u)}),(Y_1^{(v)},\ldots, Y_{N_n(v)}^{(v)})$ converges in distribution to $K=\sup_{t\in[0,1]} |B(t) |$ as $n\to\infty$, where $B=(B(t):t\in [0,1])$ is a  Brownian bridge.
Since $P(K>c_{\alpha})=\alpha$, the result follows.
\end{proof}

The Proposition \ref{prop:2} reads as follows.

\begin{proposition}
\label{prop:2}
Let $(\bar{\tau},\bar{p})$ be a  probabilistic context tree and $\{\bar{Q}^w:w\in\bar{\tau}\}$ be a family of probability distributions on $(L^{2}([0,T]), \mathcal{B}(L^{2}([0,T])))$ satisfying the assumptions of Theorem \ref{thm:2}. Let $\alpha\in (0,1)$ and define  $c_{\alpha}=(-1/2\ln(\alpha/2))^{1/2}$. For any string $s\in A^*$ such that $B_{\bar{\tau}}(s)$ is a terminal branch there exists a pair $w,w'\in B_{\bar{\tau}}(s)$
such that for almost all realization of a Brownian motion $W=(W(t):t\in[0,T])$ on $[0,T]$, 
$$
\lim_{n\to\infty}P( D_n^W((Y_1^{(w)},\ldots, Y_{N_n(w)}^{(w)}),(Y_1^{(w')},\ldots, Y_{N_n(w')}^{(w')}))\leq c_{\alpha})=0.
$$
 \end{proposition}
\begin{proof}
For each $n\geq 1$, define 
$$
N_n:=\sqrt{\frac{N_n(w)N_n(w')}{N_n(w)+N_n(w')}}, 
$$
if $\min\{N_n(w), N_n(w)\}\geq 1$.  Otherwise, we set $N_n=0$.
Observe that the strong law of large numbers implies that $P$-almost surely $N_n\to +\infty$  as $n\to\infty.$ 

Now Theorem 3.1(b) of \cite{Cuesta2006} implies that for almost all realization of a Brownian motion $W$ on $L^2([0,T])$, \begin{equation}
\label{prop.2:eq.2}
\liminf_{n\to\infty}KS(\hat{Q}^{W,w}_n,\hat{Q}^{W,w'}_n)>0 \ P\mbox{-a.s}.
\end{equation}
Since $D^W((Y_1^{(w)},\ldots, Y_{N_n(w)}^{(w)}),(Y_1^{(w')},\ldots, Y_{N_n(w')}^{(w')})=N_nKS(\hat{Q}^{h,w}_n,\hat{Q}^{h,w'}_n)$, the result follows.
\end{proof}

\begin{proof}[Proof of Theorem \ref{thm:2}]
Let $C_{\bar{\tau}}$ be the set of contexts belonging to a terminal branch of $\bar{\tau}$. 
Define also the following events
$$U_n=\bigcup_{w\in C_{\bar{\tau}}}\{\Delta^W_n(\mbox{suf}(w))\leq c\} \ \mbox{and} \ O_n=\bigcup_{w\in\bar{\tau}}\bigcup_{\substack{s\succ w:\\ \ell(s)\leq L}}\{\Delta^W_n(s)>c\}.$$
It follows from the definition of the Algorithm \ref{alg:pruning} that 
$$
P(\hat{\tau}_n\neq \bar{\tau})=P(U_n)+P(O_n).
$$
Thus, it is enough to prove that for any $\epsilon>0$ there exists $n_0=n_0(\epsilon)$ such that 
$P(U_n)\leq \epsilon/2$ and $P(O_n)\leq \epsilon/2$ for all $n\geq n_0$. 
%We start by proving that $P(U_n)\to 0$ as $n\to\infty$.

By the union bound, we see that
\begin{equation}
\label{thm.1:eq.1}
P(U_n)\leq \sum_{w\in\bar{\tau}}P(\Delta^W_n(\mbox{suf}(w))\leq c).
\end{equation} 

Since for each $w\in C_{\bar{\tau}}$, $B_{\bar{\tau}}(\mbox{suf}(w))$ is a terminal branch, Assumption \ref{ass:1} implies that there exits a pair $u,v\in B_{\bar{\tau}}(\mbox{suf}(w))$ whose associated distributions $\bar{Q}^u$ and $\bar{Q}^v$ on $F$ are different, and either $\bar{Q}^u$ or $\bar{Q}^v$ satisfy the Carleman condition.
Since $$\{\Delta^W_n(\mbox{suf}(w))\leq c\}\subset\{D^W_n((Y_1^{(u)},\ldots, Y_{N_n(u)}^{(u)}),(Y_1^{(v)},\ldots, Y_{N_n(v)}^{(v)})\leq c\},$$ Proposition \ref{prop:2} implies that $P(U_n)\to 0$ as $n\to\infty$, if $c=c_{\alpha}$ for any fixed $\alpha\in(0,1).$ As a consequence, for any $\epsilon>0$ there exists $n_0=n_0(\epsilon)$ such that $P(U_n)\leq \epsilon/2$ for all $n\geq n_0$, provided that
$c=c_{\alpha}$ with $\alpha\in(0,1)$ fixed.

Using again the union bound, we have
\begin{equation}
\label{thm.1:eq.2}
P(O_n)\leq \sum_{w\in\bar{\tau}}\sum_{\substack{s\succ w:\\ \ell(s)\leq L}}P(\Delta^W_n(s)> c).
\end{equation}
Observing that
$$
\{\Delta^W_n(s)> c\}=\bigcup_{a,b\in A}\{D^W_n((Y_1^{(as)},\ldots, Y_{N_n(as)}^{(as)}),(Y_1^{(bs)},\ldots, Y_{N_n(bs)}^{(bs)})> c\},
$$ 
we deduce from Proposition \ref{prop:1} and inequality \eqref{thm.1:eq.2} that for any $\epsilon>0$,
we can find $\alpha\in (0,1)$ sufficiently small such that if $c=c_{\alpha}$, then  $P(O_n)\leq 
\epsilon/2$ for all $n$ large enough. This concludes the proof of the theorem.
\end{proof}

\subsection{Proof of Proposition \ref{prop:bootstrap}}
\label{App:3}
\begin{proof}
 For each $a,b\in  A$ with $a\neq b$ and $1\leq i\leq N$, we set
$$Z^{a,b}_{i,n}=D^{W_i}_n\big((Y_1^{(as)},\ldots, Y_{N_n(as)}^{(as)}),(Y_1^{(bs)},\ldots, Y_{N_n(bs)}^{(bs)})\big).$$
 Assume that the null assumption $H^{(s)}_0$ is true.  
 Then the asymptotic properties of the Kolmogorov-Smirnov statistics ensure that for each $a,b\in  A$, with $a\neq b$ and $1\leq i\leq N,$ $Z^{a,b}_{i,n}$ converges in distribution to $K=\sup_{t\in[0,1]} |B(t) |$ as $n\to\infty$, where $B=(B(t):t\in [0,1])$ is a  Brownian bridge.
Since $\Delta^{W_i}_n(s)=\max_{a,b\in A:a\neq b}Z^{a,b}_{i,n}$, the continuous mapping theorem implies that
$\Delta^{W_i}_n(s)$ also converges in distribution to $K$ as $n\to\infty$, for each $1\leq i\leq N.$
In what follows,  for each $1\leq i\leq N$, we define 
$$Z_{i,n}=1\{\Delta^{W_i}_n(s)>c_{\alpha}\}.$$

To prove the Proposition, it suffices to show that for any $a_1,\ldots,a_N\in\{0,1\}$,
\begin{equation}
\label{converge_distr}
\lim_{n\to \infty}P(Z_{i,n}=a_i,\ldots, Z_{i,N}=a_N)=\alpha^{\sum_{i=1}^N a_i}(1-\alpha)^{(N-\sum_{i=1}^N a_i)}
\end{equation}

Denote $\mathcal{G}=\sigma(Y^{(as)}_k,k\geq 1,a\in A)$ and notice that conditionally on $\mathcal{G}$, the random variables $Z_{1,n}, \ldots, Z_{N,n} $ are independent for all  $n\geq 1$.
By the Skorohod's representation theorem, there is a sequence of random vectors $(\tilde{Z}_{1,n},\ldots, \tilde{Z}_{N,n})_{n\geq 1}$ and a sequence of random elements $(\tilde{Y}^{(as)}_k)_{k\geq 1,a\in A}$ taking values in $L^{2}([0,T])$, both sequences defined in the same probability space $(\tilde{\Omega},\tilde{\mathcal{F}},\tilde{P})$  such that 
\begin{enumerate}
\item  for each $n$, $(\tilde{Z}_{1,n},\ldots, \tilde{Z}_{N,n})$ has the same distribution as $(Z_{1,n},\ldots, Z_{N,n})$,
\item for each $k$ and $a\in A$, the distribution of $\tilde{Y}^{(as)}_k$ is the same as the distribution of $Y^{(as)}_k.$ 
\item if $\tilde{\mathcal{G}}=\sigma(\tilde{Y}^{(as)}_k, k\geq 1, a\in A)$, then $\tilde{Z}_{1,n},\ldots, \tilde{Z}_{N,n}$ are conditionally independent given $\tilde{\mathcal{G}}$,
\item for each $1\leq i\leq N$, $\tilde{Z}_{i,n}\to K$ almost surely with respect to $\tilde{P}$ as $n\to\infty$. 
\end{enumerate}
Item 4 and the Dominate convergence theorem for conditional expectation imply that $\tilde{P}$-a.s as $n\to\infty$, for each $1\leq i\leq N$ and $a_i\in\{0,1\}$,
$$
\tilde{P}(\tilde{Z}_{i,n}=a_i|\tilde{\mathcal{G}})\to \alpha^{a_i}(1-\alpha)^{(1-a_i)}.
$$  
Therefore, by Item 3 and the Dominate convergence theorem, we have that for any $a_1,\ldots,a_N\in\{0,1\}$, as $n\to \infty$,
\begin{eqnarray*}
\tilde{P}(\cap_{i=1}^{N}\tilde{Z}_{i,n}=a_i)&=&\tilde{E}\left[\prod_{i=1}^N\tilde{P}(\tilde{Z}_{i,n}=a_i |\tilde{\mathcal{G}})\right]\to \alpha^{\sum_{i=1}^N a_i}(1-\alpha)^{(N-\sum_{i=1}^N a_i)} 
\end{eqnarray*}
The limit in \eqref{converge_distr} now follows from Item 1. 

\end{proof}

\section*{Acknowledgements}

We thank Marcelo Queiroz  and Sebastian Hoefle for, respectively, synthesizing the acoustic stimuli and setting the experimental design. We thank also Erika Lazary and Douglas Rodrigues for helping the authros collecting the experimental data. 

This work is part of USP project {\em Mathematics, computation, language
and the brain}, FAPESP project {\em Research, Innovation and
Dissemination Center for Neuromathematics} (grant 2013/07699-0), projects {\em Stochastic modeling of the brain activity}
(CNPq grant 480108/2012-9) and {\em Plasticity in the brain after a brachial plexus lesion} (CNPq grant 478537/2012-3 and FAPERJ grants E26/110.526/2012 and E26/010.002902/2014).

\bibliography{Bibliografia}{}

\begin{thebibliography}{12}
% BibTex style file: imsart-nameyear.bst, 2013-01-28
% Default style options (sort=1,type=nameyear).
% Used options (sort=1,type=nameyear).

\bibitem[\protect\citeauthoryear{B\"uhlmann and Wyner}{1999}]{buhlmann99}
\begin{barticle}[author]
\bauthor{\bsnm{B\"uhlmann},~\bfnm{Peter}\binits{P.}} \AND
  \bauthor{\bsnm{Wyner},~\bfnm{Abraham~J.}\binits{A.~J.}}
(\byear{1999}).
\btitle{Variable length Markov chains}.
\bjournal{Ann. Statist.}
\bvolume{27}
\bpages{480--513}.
\bdoi{10.1214/aos/1018031204}
\end{barticle}
\endbibitem

\bibitem[\protect\citeauthoryear{Cuesta-Albertos, Fraiman and
  Ransford}{2006}]{Cuesta2006}
\begin{barticle}[author]
\bauthor{\bsnm{Cuesta-Albertos},~\bfnm{J.~A.}\binits{J.~A.}},
  \bauthor{\bsnm{Fraiman},~\bfnm{R.}\binits{R.}} \AND
  \bauthor{\bsnm{Ransford},~\bfnm{T.}\binits{T.}}
(\byear{2006}).
\btitle{Random projections and goodness-of-fit tests in infinite-dimensional
  spaces}.
\bjournal{Bulletin of the Brazilian Mathematical Society, New Series}
\bvolume{37}
\bpages{477-501}.
\bdoi{10.1007/s00574-006-0023-0}
\end{barticle}
\endbibitem

\bibitem[\protect\citeauthoryear{Cuesta-Albertos et~al.}{2007}]{Cuestaetal:07}
\begin{barticle}[author]
\bauthor{\bsnm{Cuesta-Albertos},~\bfnm{Juan~Antonio}\binits{J.~A.}},
  \bauthor{\bsnm{Fraiman},~\bfnm{Ricardo}\binits{R.}},
  \bauthor{\bsnm{Galves},~\bfnm{Antonio}\binits{A.}},
  \bauthor{\bsnm{Garcia},~\bfnm{J.~E.}\binits{J.~E.}} \AND
  \bauthor{\bsnm{Svarc},~\bfnm{Marcela}\binits{M.}}
(\byear{2007}).
\btitle{Classifying Speech Sonority Functional Data using a Projected
  Kolmogorov-Smirnov Approach}.
\bjournal{Journal of Applied Statistics}
\bvolume{34}
\bpages{749-761}.
\bdoi{10.1080/02664760701237077}
\end{barticle}
\endbibitem

\bibitem[\protect\citeauthoryear{Delorme and Makeig}{2004}]{Delorme-eeglab:04}
\begin{barticle}[author]
\bauthor{\bsnm{Delorme},~\bfnm{A.}\binits{A.}} \AND
  \bauthor{\bsnm{Makeig},~\bfnm{S.}\binits{S.}}
(\byear{2004}).
\btitle{EEGLAB: an open source toolbox for analysis of single-trial EEG
  dynamics including independent component analysis}.
\bjournal{Journal of Neuroscience Methods}
\bvolume{134}
\bpages{9--21}.
\end{barticle}
\endbibitem

\bibitem[\protect\citeauthoryear{Galves et~al.}{2012}]{galves:12}
\begin{barticle}[author]
\bauthor{\bsnm{Galves},~\bfnm{Antonio}\binits{A.}},
  \bauthor{\bsnm{Galves},~\bfnm{Charlotte}\binits{C.}}, \bauthor{\bsnm{Garc{\'
  \i}~a},~\bfnm{J.~E.}\binits{J.~E.}},
  \bauthor{\bsnm{Garcia},~\bfnm{Nancy~L.}\binits{N.~L.}} \AND
  \bauthor{\bsnm{Leonardi},~\bfnm{Florencia}\binits{F.}}
(\byear{2012}).
\btitle{Context tree selection and linguistic rhythm retrieval from written
  texts.}
\bjournal{Ann. Appl. Stat.}
\bvolume{6}
\bpages{186--209}.
\bdoi{10.1214/11-AOAS511}
\end{barticle}
\endbibitem

\bibitem[\protect\citeauthoryear{Garrido, Sahani and Dolan}{2013}]{Garrido:13}
\begin{barticle}[author]
\bauthor{\bsnm{Garrido},~\bfnm{Marta~I.}\binits{M.~I.}},
  \bauthor{\bsnm{Sahani},~\bfnm{Maneesh}\binits{M.}} \AND
  \bauthor{\bsnm{Dolan},~\bfnm{Raymond~J.}\binits{R.~J.}}
(\byear{2013}).
\btitle{Outlier responses reflect sensitivity to statistical structure in the
  human brain.}
\bjournal{PLOS Computational Biology}
\bvolume{9}.
\bdoi{doi:10.1371/journal.pcbi.1002999}
\end{barticle}
\endbibitem

\bibitem[\protect\citeauthoryear{N\"a\"at\"anen, Gaillard and
  M\"antysalo}{1978}]{Naatanen:78}
\begin{barticle}[author]
\bauthor{\bsnm{N\"a\"at\"anen},~\bfnm{R.}\binits{R.}},
  \bauthor{\bsnm{Gaillard},~\bfnm{A.~W.~K.}\binits{A.~W.~K.}} \AND
  \bauthor{\bsnm{M\"antysalo},~\bfnm{S.}\binits{S.}}
(\byear{1978}).
\btitle{Early selective-attention effect on evoked potential reinterpreted}.
\bjournal{Acta Psychologica}
\bvolume{42}
\bpages{313 - 329}.
\bdoi{https://doi.org/10.1016/0001-6918(78)90006-9}
\end{barticle}
\endbibitem

\bibitem[\protect\citeauthoryear{N\"a\"at\"anen, Jacobsen and
  Winkler}{2005}]{Naatanem:2005}
\begin{barticle}[author]
\bauthor{\bsnm{N\"a\"at\"anen},~\bfnm{Risto}\binits{R.}},
  \bauthor{\bsnm{Jacobsen},~\bfnm{Thomas}\binits{T.}} \AND
  \bauthor{\bsnm{Winkler},~\bfnm{Istvan}\binits{I.}}
(\byear{2005}).
\btitle{Memory-based or afferent processes in mismatch negativity (MMN): A
  review of the evidence}.
\bjournal{Psychophysiology}
\bvolume{42}
\bpages{25--32}.
\bdoi{10.1111/j.1469-8986.2005.00256.x}
\end{barticle}
\endbibitem

\bibitem[\protect\citeauthoryear{Rabiner and Juang}{1993}]{Rabiner:1993}
\begin{bbook}[author]
\bauthor{\bsnm{Rabiner},~\bfnm{Lawrence}\binits{L.}} \AND
  \bauthor{\bsnm{Juang},~\bfnm{Biing-Hwang}\binits{B.-H.}}
(\byear{1993}).
\btitle{Fundamentals of Speech Recognition}.
\bpublisher{Prentice-Hall, Inc.}, \baddress{Upper Saddle River, NJ, USA}.
\end{bbook}
\endbibitem

\bibitem[\protect\citeauthoryear{Rissanen}{1983}]{Rissanen:83}
\begin{barticle}[author]
\bauthor{\bsnm{Rissanen},~\bfnm{J.}\binits{J.}}
(\byear{1983}).
\btitle{A Universal Data Compression System}.
\bjournal{IEEE Trans. Inf. Theor.}
\bvolume{29}
\bpages{656--664}.
\end{barticle}
\endbibitem

\bibitem[\protect\citeauthoryear{von Helmholtz}{1867}]{VonHelmholtz:67}
\begin{bbook}[author]
\bauthor{\bparticle{von} \bsnm{Helmholtz},~\bfnm{H.}\binits{H.}}
(\byear{1867}).
\btitle{Handbuch der physiologischen Optik}
\bvolume{III}.
\bpublisher{Leopold Voss}
\bnote{translated by The Optical Society of America in 1924 from the third
  germand edition, 1910, Treatise on physiological optics, Vol. III}.
\end{bbook}
\endbibitem

\bibitem[\protect\citeauthoryear{Wacongne, Changeux and
  Dehaene}{2012}]{Wacongne:12}
\begin{barticle}[author]
\bauthor{\bsnm{Wacongne},~\bfnm{C.}\binits{C.}},
  \bauthor{\bsnm{Changeux},~\bfnm{J.~P.}\binits{J.~P.}} \AND
  \bauthor{\bsnm{Dehaene},~\bfnm{S.}\binits{S.}}
(\byear{2012}).
\btitle{A Neuronal Model of Predictive Coding Accounting for the Mismatch
  Negativity}.
\bjournal{The Journal of Neuroscience}
\bvolume{32}
\bpages{3665-3678}.
\bdoi{10.1523/JNEUROSCI.5003-11.2012}
\end{barticle}
\endbibitem

\end{thebibliography}
\bibliographystyle{imsart-nameyear}
\nocite{Rissanen:83}
\nocite{VonHelmholtz:67}
\nocite{Delorme-eeglab:04}
\nocite{Cuesta2006}
\nocite{Wacongne:12}
\nocite{buhlmann99}
\nocite{Garrido:13}
\nocite{galves:12}
\nocite{Naatanem:2005}
\nocite{Naatanen:78}
\nocite{Rabiner:1993}
\end{document}